# Multi-class Twitter Data Categorization and Geocoding with a Novel Computing Framework

**Sakib Mahmud Khan[1]\*, Mashrur Chowdhury[2], Linh B. Ngo [3], Amy Apon [4]**

[1] PhD Candidate, Glenn Department of Civil Engineering, Clemson University, Clemson, SC 29634, USA; email: sakibk@g.clemson.edu

[2] PhD, Professor, Glenn Department of Civil Engineering, Clemson University, Clemson, SC 29634, USA; email: mac@ clemson.edu

[3] PhD, Director of Data Science, Cyberinfrastructure and Technology Integration, Clemson University, SC 29634, USA; email: lngo@clemson.edu

[4] PhD, Professor, Chair of Computer Science Division, School of Computing, Clemson University, Clemson, SC 29634, USA; email: aapon@clemson.edu

\*Corresponding author

## Abstract

This study details the progress in transportation data analysis with a novel computing framework in keeping with the continuous evolution of the computing technology. The computing framework combines the Labelled Latent Dirichlet Allocation (L-LDA)-incorporated Support Vector Machine (SVM) classifier with the supporting computing strategy on publicly available Twitter data in determining transportation-related events to provide reliable information to travelers. The analytical approach includes analyzing tweets using text classification and geocoding locations based on string similarity. A case study conducted for the New York City and its surrounding areas demonstrates the feasibility of the analytical approach. Approximately 700,010 tweets are analyzed to extract relevant transportation-related information for one week. The SVM classifier achieves more than 85% accuracy in identifying transportation-related tweets from structured data. To further categorize the transportation-related tweets into sub-classes: incident, congestion, construction, special events, and other events, three supervised classifiers are used: L-LDA, SVM, and L-LDA incorporated SVM. Findings from this study demonstrate that the analytical framework, which uses the L-LDA incorporated SVM, can classify roadway transportation-related data from Twitter with over 98.3% accuracy, which is significantly higher than the accuracies achieved by standalone L-LDA and SVM.

## 1. Introduction

Traffic information is currently available through different private sources and navigation applications developed by private companies, such as Waze, Google, or Apple. At the same time, public agencies, specifically law enforcement agencies, must collect, validate, and disseminate incident information, as they are primarily responsible for traffic management and safety. A 2015 survey found that most state transportation agencies collect traffic data from sensors and through third parties, such as INRIX, and then use web sites and Dynamic Message Signs to disseminate traffic information to travelers (Fries et al., 2015). In the study conducted by Fries et al. (2015), based on the survey responses, researchers emphasized the need for improvement in methods and technologies for travel time data collection. As stated in a USDOT (2018) report, transportation applications using real-time data increases the operational and safety benefits by generating data helpful for making informed travel decisions (USDOT, 2018). Given the importance of the quality and availability of traffic data for providing reliable transportation services, tools that provide accurate, timely and accessible data to support traffic management and planning practices related to traffic information dissemination are essential. In addition to navigation applications developed by private companies, social media platforms like Twitter produce publicly available data that can provide 'where', 'what' and 'when' information about any traffic incident event. For example, "Incident on #MontaukBranch EB at Jamaica Station" tweet says where (i.e., at MontaukBranch EB, Jamaica Station) and what event (i.e., incident) happened. Another example tweet, "real confused as to why the workers aren't out here cleaning the roads!!" tells what event (i.e., there are obstructions or debris on the road), but the tweet itself does not tell where the event happened unless tweet has geolocation information available beyond the tweet text. In both tweet examples, the time of tweet generation is provided by Twitter. While Twitter has been analyzed as a potential source of traffic data (D'Andrea, Ducange, Lazzerini, & Marcelloni, 2015; Gu, Qian, & Chen, 2016), tweets do not always have geolocation information available. Also, since drivers should not tweet while driving, Twitter data is most appropriate as support for traffic incident-related data in which the tweets from the general public originate from stopped vehicles or the passengers within (Pratt, Morris, Zhou, Khan, & Chowdhury, 2019).

In this paper, the term 'tweet' refers to the message or status update from a Twitter user account, which cannot exceed the 140 character limit (the size of tweets has been extended to 280 characters since the time of this study). Although Twitter provides data generated by numerous users from a specific region, analyzing the raw streaming data in real-time and providing useful feedback based on the analysis are challenging. The research objective is to develop a parallel-computing based analytical framework to accurately categorize and reliably geocode tweets for the transportation-related events. This contribution of this paper entails developing and evaluating: (a) the Labelled Latent Dirichlet Allocation (L-LDA)-incorporated Support Vector Machine (SVM) classifier to classify tweets with supporting distributed computing framework to support roadway transportation operations and (b) the string-similarity based location identification system.

After analyzing the collected tweets from a specific region using the Natural Language Processing (NLP) techniques, transportation-related tweets are extracted with SVM, a supervised classification technique. SVM is used to identify transportation-related tweets from the whole Twitter dataset for each day, and the Clemson University Palmetto supercomputing cluster is used to support parallel computations to develop SVM models. The motivation of using this parallel computation framework, to classify almost 700,010 tweets in this study, is to minimize the computation time for the SVM training phase compared to single node-based computation. After identification, the transportation-related tweets are classified via three supervised classification techniques: L-LDA, SVM, and L-LDA incorporated SVM. L-LDA is a supervised credit attribution method, whereas L-LDA and L-LDA incorporated SVM have not been used to identify transportation-related events in earlier research. It has been previously determined that L-LDA performs as well as or better than SVM for multi-label text classification (Ramage, Hall, Nallapati, & Manning, 2009). The motivation for integrating L-LDA with SVM in this study is to improve the performance of SVM in classifying tweets. In the L-LDA incorporated SVM technique, topic distribution probability for each tweet generated by L-LDA is used by SVM classifier to categorize the tweets in multiple classes (i.e., incident, congestion, special event, construction, and other events). Accuracies of SVM, L-LDA, and L-LDA incorporated SVM classifiers are measured with respect to the labels manually assigned to the tweets.

According to Title 23 of the Code of Federal Regulations, real-time highway information programs, including statewide incident reporting system, must be 85% accurate as a minimum (GPO, 2011). It can be inferred, from this code, that it is possible to use Twitter as a potential standalone tool to compile and classify roadway transportation events if the accuracy is above the 85% threshold. Following the text classification, the tweets are geocoded. Using the analytical framework presented in this study, a case study is conducted for New York City (NYC) and its surrounding areas. The following sections discuss the previous studies related to twitter data analysis, analytical framework for this study, and a case study using the analytical framework.

## 2. Literature review

Twitter data are used for assessing various events (D'Andrea et al., 2015; Gu et al., 2016; He, Boas, Mol, & Lu, 2017; Purohit et al., 2014; Qian, 2016; Roberts et al., 2018; Tang et al., 2017) including natural disasters, mass emergency, acts of terrorism, extreme weather events, political protests, and transportation events. In a study conducted by Mirończuk and Protasiewicz (2018), the authors have reviewed recent research to understand the general approach of text classification practices and identify the future research questions related to text classification (Mirończuk & Protasiewicz, 2018). The most common research for text classification includes the use of supervised learning methods and involves a number of steps including data acquisition, data labeling, feature construction, feature weighing, feature selection, classification model training, and assessment. The authors have identified overfitting of the text classification models, dynamic classifier selection, multi-lingual text analysis, text stream analysis, sentiment analysis and ensemble-learning methods as the emerging research topics in text classification.

### 2.1 Tweet classification with machine learning

Once Twitter data are collected, their contents are analyzed. This is a difficult process, as Twitter data is often characterized as "vast, noisy, distributed, unstructured, and dynamic" (Gundecha & Liu, 2014). Therefore, machine-learning techniques are integral to the process of mining content for decision-making purposes. These machine learning techniques are categorized into three primary areas, supervised (Kotsiantis, 2007), semi-supervised (Zhu, 2006), and unsupervised (Hastie, Friedman, & Tibshirani, 2001). A supervised learning algorithm uses training data with known outcomes. The learning algorithm can gradually adjust its parameters to generate results from training data so that these results match most closely with the known outcomes. For unsupervised learning, there are no known outcomes, and the algorithm will attempt to extract the pattern from the data itself. Semi-supervised learning techniques contain a mixture of both by using a small set of training data with known outcomes and a majority of training data without known outcomes. The evidence of the various degrees of success in applying different machine learning techniques to analyze social media contents is well known (D'Andrea et al., 2015; Ramage et al., 2009). For this specific study, a supervised machine learning technique, SVM (Cortes & Vapnik, 1995), is selected to automate the process of identifying transportation/non-transportation tweets, and L-LDA (Ramage et al., 2009), another supervised technique, is selected to model the topics of the classified tweets. SVM facilitates the utilization of kernel functions to develop hyperplane(s) within the feature space of the observation to classify the observations into different distinctive groups. Supervised LDA (s-LDA) methods are used to identify the label of the tweets by simply constraining the topic model to use only the topics corresponding to the training dataset's label set. s-LDA is used by Gu et al. (2016), where the authors found that 51% of the geo-codable tweets can A similar with the s-LDA classifier (Gu et al., 2016).

### 2.2 Twitter data for transportation applications

In earlier investigations of the reliability and accuracy of social media data for unplanned transportation events (i.e., incidents, congestion), various methods (i.e., machine learning, statistical analysis) were proposed to extract necessary data from user-focused contextual information that is shared in the social media platform. To determine real-time incident information, Twitter data were analyzed using machine learning technique that incorporated semantic web technology (i.e., Linked Open Data Cloud) and features from tweets and LOD data for tweet classification (i.e., car crash class, shooting class and fire class) (Schulz & Ristoski, 2013). The proposed model achieved about 89% accuracy for classifying tweets. They concluded that even with very few social media posts, this method is capable of detecting incidents. For traffic congestion monitoring, (Chen, Chen, & Qian, 2014) developed a statistical framework that integrated both Hinge-loss Markov Random Fields and a language model. Evaluations were performed over different spatial-temporal and other performance metrics on the collected tweet and INRIX probe datasets. The two major U.S. cities used in this study were Washington D.C. and Philadelphia, PA. Based on their analysis, (Chen et al., 2014) found that Twitter data can supplement traditional road sensor data to assess traffic operational conditions. The authors from (Sakaki, Matsuo, Yanagihara, Chandrasiri, & Nawa, 2012) study created a system to distribute important event-related information to vehicle drivers, including the location information and temporal information. Tweets were classified as either traffic or not-traffic related. Subsequently, the extracted information was forwarded to vehicle drivers after extracting spatial information from the tweets. As a result, the authors achieved an 87% precision rate in categorizing tweets that referred to heavy traffic. To classify incident-related tweets, Gu et al. (2016) utilized an adaptive data acquisition framework and prepared a dictionary of important keywords. The study suggested that the mining of Twitter data holds potential to cost-effectively providing traffic incident data. Additional findings noted that most of the geo-tagged tweets are posted by influential users who are mainly public agencies or/and media (Gu et al., 2016). In their comparison of Twitter data analysis for road incident events from the California Highway Patrol (CHP), The authors from (Mai & Hranac, 2013) study captured tweet based on specific keywords. The authors used a nine-hour time window and a 50-mile radius to match tweets with CHP records and then applied a semantic-based weighting factor. They suggested a logical order of Twitter analysis, which involves identifying tweets with correctly geocoded information (latitude and longitude), filtering tweets that contain traffic information, and analyzing these tweets. The approach is limited in that only a small percentage of tweets contained latitude and longitude (Gu et al., 2016). Compared to the complete data set acquired from Twitter's Firehose, it is possible to infer that the number of usable tweets is further reduced in cases where Twitter's public API is used, due to the 1% of the total data available in the public API.

### 2.3 Twitter data analysis in a distributed computing infrastructure

Large-scale data analysis in a centralized environment is often inefficient, and impractical due to the high computation time. As such, applications of parallel computing framework in civil engineering decision making have been developed in (Kandil & El-Rayes, 2005; Karatas & El-Rayes, 2015). In (Kandil & El-Rayes, 2005) the authors used a manager/worker paradigm and a distributed genetic algorithm to optimize both the construction time and costs of large-scale construction projects. The input of the optimization tasks were project planning data that described project activities. Inititally, the processor functioning as a manager in the manager/worker paradigm initialized a genetic algorithm to create a random set of feasible solutions. Finally, the manager processor completed the fitness evaluation to generate a new set of solutions. Using 150 experiments on the parallel computin cluster at the University of Illinois, the authors found an eight-time parallel speedup in obtaining solutions compared to the single processing framework. Similarly, the authors in (Karatas & El-Rayes, 2015) evaluated a parallel computation-enabled genetic algorithm where multiple processors analyzed the environmental impacts of a subpopulation distributed by the coordinator processor. Based on the fitness function evaluation from the multiple processors, the coordinator processor creates the next group of solutions. The computation time was reduced to 1.7 days from 12 days using eight paralleled processors. The distributed computing framework was also studied in terms of analyzing large-scale Twitter data. The authors in (Gao, Ferrara, & Qiu, 2015) studied parallel clustering of social media data using the stream processing engine, Apache Storm, which helps to implement parallel processors and distribute workload in a fault tolerant environment. First, the initial clusters were developed using historical Twitter data. Based on these initial clusters, multiple processors clustered the new tweet stream and detected outliers. Using the framework, the computation speed with 96 parallel processors was higher than the Twitter stream arrival speed. The authors in (Kanavos et al., 2017) used MapReduce and Apache Spark framework to classify tweet sentiments based on hashtag and emoticons. With the increase in data size, the analysis speed increased linearly with the increase in processor number. A similar study on tweet sentiment analysis conducted in (Kumar & Rahman, 2017) entailed evaluating the Apache Spark and Message Passing Interface (MPI) clustering frameworks. MPI performed better than Apache Spark in that the programmer had access to the freedom-on-memory allocation and task scheduling. So far, no study has been conducted on distributing supervised machine learning methods to classify transportation-related tweets, which is a motivation for this study.

## 3. L-LDA incorporated SVM

The L-LDA incorporated SVM classifier is a supervised learning based classifier, in which the feature space of the SVM includes the multinomial topic distributions ($\theta$) value over the vocabulary for each topic generated by L-LDA. The L-LDA classifies the tweets based on the mixture of the underlying topic. The main difference between traditional LDA (Blei, Ng, & Jordan, 2003) and L-LDA (Ramage et al., 2009) is that L-LDA constrains the topic model to use topics observed in a training data set. For a processed tweet $T$, let us consider $N$ is the total vocabulary size in $T$, expressed as a tuple $\mathbf{w} = (w_1, ..., w_i, ... w_N)$, where $w_i$ is an $i$-th processed token. Each $\mathbf{w}$ is accompanied with a label presence/absence indicator list $\mathbf{L} = (l_1, l_2, ..., l_K)$ where $l_i = 1$ is the topic i presence indicator and $l_i = 0$ is the topic i absence indicator. There are $K$ topics in the training set. The multinomial mixture distribution $(\theta)$ is used to identify the final label of the test data, which is restricted to only topics $K$ from the training dataset, meaning L-LDA assigns a label for a test case based on the training dataset label.

The feature space is the main difference between the L-LDA incorporated SVM and the SVM classifier. Using L-LDA, the multinomial topic distributions ($\theta$) values of each tweet is estimated, and these topic distribution values are identified by (Mirończuk & Protasiewicz, 2018). As detailed in the analytical approach in Fig. 1, tweets are first collected from a specific region. The data processing, feature extraction, feature selection, and classification steps are then associated with tweet classification to identify the relevant transportation-related tweets. Based on the data size and computation complexities, parallel computation is used to increase the data processing and tweet classification capabilities of the included in both training and test feature space of the L-LDA incorporated SVM. Cross-validation is used to identify the transportation sub-class specific $\theta$ values, which gives better performance compared to the standalone SVM. The SVM classifier, as considered in this study, lacks the support on the topic distribution, unlike the L-LDA incorporated SVM.

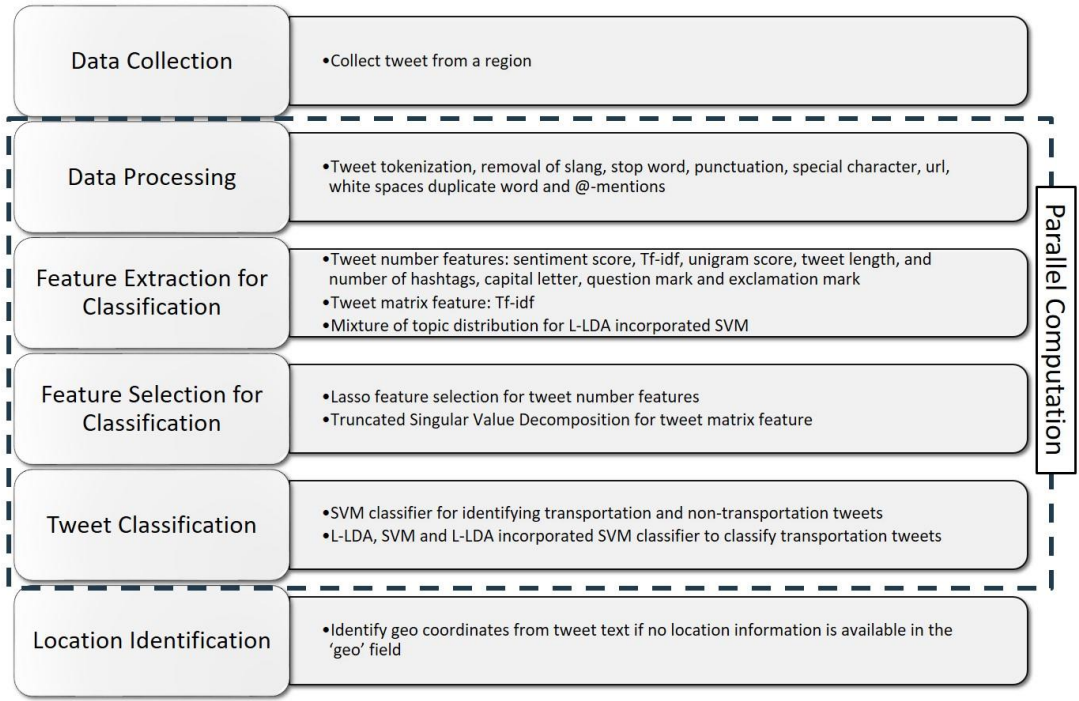

**Fig. 1.** Analytical approach steps.

## 4. An analytical approach for twitter categorizing and geocoding

This research utilizes the same procedure as described by (Mirończuk & Protasiewicz, 2018) to develop a text stream analysis framework for a large region. The motivation of this research is to satisfy the research gap in text stream analysis as framework. Next, location information is extracted from the tweet from 'geo' field, and if no coordinate is available, the location information is extracted from the tweet text.

A case study is conducted for NYC and its surrounding areas using the adopted analytical framework. Tweets are collected for the week of Saturday, 01/07/2017 to Friday 01/13/2017). First, data for two days of the week are labeled: Saturday and Wednesday with the total volume of the generated tweet for these two days 194K. Five individuals have helped to label these data. The total rate of manual labeling is almost 3000 tweets/hour. When the ground truth data is labeled, the SVM supervised classifier is developed based on these two days data. In the second step, the SVM classifier is used to classify the data for the remaining five days. However, given the poor performance of the supervised classifier in classifying the unstructured data, manual labeling is conducted based on a keyword search for the other five days. The keywords are selected based on the data from Saturday and Wednesday, and also from other literature. These tweets are categorized manually into transportation-related tweets. The SVM classifiers accuracy, precision, recall, and Root Mean Square Error (RMSE) are subsequently studied to categorize non-transportation related and transportation-related tweets. Following one annotator-one manager approach, data are divided into different parts, with each passed to each individual annotator. The work of the annotator was verified by the manager to create the ground truth data. The parallel computation nodes on the Palmetto Supercomputing cluster at Clemson University are then used to develop and evaluate the classifier. After studying the accuracy of SVM in identifying the transportation-related tweets, the accuracies of supervised L-LDA, SVM and L-LDA incorporated SVM are investigated to identify five sub-classes (i.e., construction, traffic operations, incidents, special events, and other events). The tweets are then passed through two geocoders to determine the tweet location. The steps of the analytical approach are described below.

### 4.1 Data collection

Using the Twitter streaming API, tweets from NYC and its surrounding areas, confined by approximately (40.49, -74.25) and (40.92, -73.70) coordinates, are collected using a location-bounding box which covered all five boroughs (i.e., county-level administrative divisions) of NYC and its surrounding areas. No additional features or keywords are used to collect the tweets. The total number of tweets collected for each day from Saturday to Friday are 79,310, 99,879, 106,520, 98,932, 115,391, 99,671, and 97,976, respectively. These tweets are all labeled manually to validate the accuracy of the SVM, L-LDA, and L-LDA incorporated SVM classifiers. Several students were recruited to label the tweets, and the later accuracy of the classifiers are evaluated compared to the labels assigned by the students.

### 4.2 Data preprocessing for classification

For Twitter, the streaming API returns additional information such as user id, profile information, and creation time along with the tweet text. Only tweet texts are considered for classification. Given the inherent ambiguity of tweets (e.g., non-standard spelling, inconsistent punctuation and/or capitalization), the following preprocessing steps are performed to extract the features for the classification:

- In the first step, the tweets are tokenized, meaning that they are transformed into a group of meaningful processing units (e.g., phrases, syllables, or words). Each tweet $T$ is split into words, $w$, after which each tokenized tweet $T$ is expressed as:

$$w = \{w_1, w_2, w_3, \dots, w_i, \dots, w_N\} \quad (1)$$

  where $w_i$ is the i-th tokenized word for each tweet $T$ of length $N$.

- In the second step, internet slangs are replaced and stop words are removed. Internet slangs are highly informal words, and abbreviations or expressions used by the general public for online interaction. Such slang is not considered as part of the standard language, which requires their replacement with elaborated expressions. For example, 'hbd' is replaced with 'happy birthday', and '2moro' is replaced with 'tomorrow.' Stop-words (i.e., articles, prepositions, conjunctions) are those words within a sentence that offer negligible or no information for the text analysis. In this paper, a list including both slang words (a total of 5188 records) and stop words (a total of 675 records) are created with the lists available from multiple online resources.

- In the third and final step, punctuation marks, special characters (e.g., ^, $, ., |, *, +) and additional white spaces in each tweet are removed, followed by the removal of duplicate words, and replacing the URL with the term 'URL' and @-mentions with 'at_user'.

After this processing, a tweet is expressed as a sequence of relevant tokens that excludes the stop words, punctuation marks, special characters, and duplicate tokens. If $r$ is the processed relevant token, the processed tweet $T$ is expressed as:

$$r = \{r_1, r_2, r_3, \dots, r_i, \dots, r_M\} \quad (2)$$

where $r_i$ is the $i$-th processed relevant token of processed tweet $T$ of length $M$ (excluding the stop words, punctuation marks, special characters, and duplicate tokens). $M \leq N$, where $N$ is the total token number (including the stop words, punctuation marks, special characters, and duplicate tokens) for each tweet.

### 4.3 Tweet feature extraction

Extracting features from textual data to identify the most relevant transportation-related tweets involves a conversion of tweets' texts to numeric matrices. It was determined from an earlier study (Schulz, Guckelsberger, & Schmidt, 2015) that for generalized models, (i.e., models applicable in multiple areas, that even if the training dataset is developed using data from a single area or few areas) a limited number of features containing word-n-grams and character-n-grams exhibited superior performance over a similar dataset with a large number of features. For the developed model, several unique numeric features and one tf-idf Vector are considered for the classification analysis as followed.

- Sentiment score is considered as one of the features, as the general public expresses emotions through tweet texts while traveling and/or during unplanned events (e.g., warning during congestions, incidents which will have negative sentiment values). Here, a lexicon-based analysis is performed, in which a dictionary of words with emotional connotation strength is used to measure the sentiment related with each tweet. The value of the emotional connotation expresses the polarity (i.e., positivity or negativity) the words. If a processed tweet $T$ has tokens $r = \{r_1, r_2, r_3, \ldots, r_i, \ldots, r_M\}$, then the polarity of $T$ is calculated as (Dayalani and Patil, 2014):

$$Polarity\ (\mathbf{r}) = {\sum_{i=1}^{M} P(r_i)}\Big/{N} \quad (3)$$

Where $N$ is the total token numbers in each tweet, $P(r_i)$ is the polarity score of token $r_i$ calculated from the used lexicon.

- Term Frequency Inverse Document Frequency (tf-idf) values amplify the effect of unique words for each document or single tweet and diminish the effect of common words in the whole tweet dataset or corpus because the common words contain no extra information. This feature has been used in previous studies to classify transportation-related tweets (Khatri, 2018; Schulz & Ristoski, 2013). For each processed tweet token $r$ the $idf$ is calculated, based on training corpus $D$. Following is the equation of calculating tf-idf.

$$tf\text{-}idf(r, d, D) = tf(r, d) \times idf(r, D) \quad (4)$$

where $r$ is a processed relevant token from tweet $d$ and $D$ is a corpus of tweets; $tf(r, d)$ is a frequency of $r$ in $d$ and $idf(r, D)$ is an inverse document frequency of $r$: $idf(r, D) = log({|D|}\big/{1 + |df(r,D)|})$. Here $df(r, D)$ is a number of tweets from $D$ in which $r$ occurs at least once, and $|D|$ is the total tweet number in the document. '$|x|$' represents the count of variable $x$.

- The presence of a specific word/token can help to determine the tweet category with the 'Frequent Token Presence' or FTP score calculated based on the presence of a specific word from a list in the specific tweet dataset. The list is created based on the most frequent words in the training dataset. Consider a processed tweet $T$ with token set $\mathbf{r} = \{r_1, r_2, r_3, \ldots, r_i, \ldots, r_M\}$. If $|r|$ is the total count of a token $r$ if it exists in the most frequent word list ($L$), the FTP score of tweet $T$, expressed as $FTP(\mathbf{r})$, is calculated as:

$$FTP(\mathbf{r}) = {\sum_{m=1}^{M} |r_m \subset L|}\Big/{N} \quad (5)$$

where $M$ is the total number of processed relevant tokens (excluding the stop words, punctuation marks, special characters, and duplicate tokens) in $T$.

- Syntactic features, i.e., the number of hashtags, question marks, exclamation marks, the number of capital letters, and the tweet length, are also considered.

### 4.4 Tweet feature selection

After the initial features are extracted, the relevant features required to develop the reliable classification models are selected based upon Lasso feature selection as the data may not be normally distributed (Fonti & Belitser, 2017). For tf-idf vector, a different feature selection strategy is used. For high-dimensional data like the tf-idf vector, Singular Value Decomposition (SVD) identifies the dominant pattern inside the main data. SVD maps the high-dimensional data into a new coordinate system using the correlations between the initial data. Considering a rectangular matrix $M$, SVD decomposes $M$ into three matrices as shown below.

$$M = A\ S\ B^T \qquad (6)$$

where S is a diagonal matrix, and A and B are two orthogonal matrices. A Truncated SVD or T-SVD discards the small singular values of $M$. Using T-SVD, a matrix $M_j$ with reduced rank $j$ can represent the matrix $M$ fairly accurately, which can be used for feature dimension reduction

### 4.5 Tweet classification

The selected features are standardized (i.e., the distribution of each attribute is shifted to mean of '0' and standard deviation of '1') and normalized (i.e., the numeric attributes are rescaled into the range of 0 to 1). Once all the features are normalized, the SVM classifier is used. SVM can process data with high dimensional feature spaces and a sparse document vector (Joachims, 1998). The model is implemented using Scikit-learn libraries. For the multi-class SVM problem, the one-vs-one decomposition process is used. This process handles an *'n'* class-based classification problem with *n(n-1)/2* number of binary classifiers that distinguish between different pairs of classes. The final class is assigned based on majority voting that is assigned by *n(n-1)/2* binary classifiers. Next, different kernel functions (i.e., linear, polynomial, and radial basis functions) are tested and their associated parameters are identified from cross-validation within the training dataset. For statistical confidence, 30 processes are executed concurrently on the Palmetto Supercomputer at Clemson University, as required by the non-parametric Wilcoxon Signed Rank test ("Scipy," 2019). This test is used in this research to compare the performance of supervised classifiers. A PBS script is written to run all the test cases in parallel. The requested interactive jobs are submitted for all test cases running simultaneously, with each using a single hardware node with 16 CPU cores per node, and 60 GB of RAM per node.

Once transportation-related tweets are identified, L-LDA, SVM, and L-LDA incorporated SVM are used to classify the transportation-related tweets in the following five topics:

- Construction: Updated status related to construction;

- Traffic operations: Updated status related to traffic;

- Incidents: Incident notification, and clearance information;

- Special events: Road closure due to the public gathering;

- Other Events: Events that do not fall under any specific category.

For L-LDA classifier, the Collapsed Variational Bayesian method is used for inference of the training model over test dataset (Teh, Newman, & Welling, 2007). For each run, the dataset is randomly divided into two groups. The initial 80% data in each run is considered as training dataset and the rest 20% of the dataset is considered as test data. The accuracy, for all classifiers, is then derived using Eq. 7 expressed as

$$Accuracy = {CT_P}/{TT_P} * 100 \ (Eq.\ 7)$$

where $CT_P$ is the correctly classified tweets, and $TT_P$ is the total tweet number. Also, precision (%) is calculated as

$$Precision = {True\ Positive}/{(True\ Positive + False\ Positive)} * 100 \qquad (8)$$

Recall (%) is calculated as

$$Recall = {True\ Positive}/{(True\ Positive + False\ Negative)} * 100 \qquad (9)$$

While computing overall recall and precision, the macro average measure is used, which shows the average recall or precision values over the total number of classes. For $C$ total class number, the macro-average value of recall can be calculated using Eq. 10. Similarly, the macro-average value of precision can be calculated.

$$Recall_{Macro-average} = {\sum_{i=1}^{C} Recall_i}/{C} \qquad (10)$$

With sample size $A$ if $y_{obs,i}$ is the observed $i$-th data and $y_{for,i}$ is the forecasted $i$-th data, RMSE can be calculated with the following Eq. 11.

$$RMSE = \sqrt{({\sum_{i=1}^{A}(y_{for,i} - y_{obs,i})^2}/{A})} \qquad (11)$$

Shapiro-Wilk and Anderson-Darling univariate normality tests are used to check whether the underlying data is normally distributed or not. As the underlying data are not normally distributed, the non-parametric statistical test, Wilcoxon Signed Rank test is used to compare the median of paired samples. As same test datasets are used to evaluate L-LDA incorporated SVM, SVM and L-LDA classifiers, the paired sample test, i.e., Wilcoxon Signed Rank test is used. The hypotheses (Stephanie, 2005) are as follows:

$H_0$ = the medians of classifier accuracies are equal
$H_A$ = the medians of classifier accuracies are not equal

If 0.1 level of significance is considered, then the $H_0$ (i.e., null hypothesis) is rejected when p-values < 0.1.

Loper, 2009). The NER is used to capture street information via the following steps:

1. From the original tweet, @, URL, and hashtag signs are removed, and hyphen sign was replaced with 'or'. After this processing task, tokens for each tweet are extracted.
2. Using the tokens, the Part-Of-Speech (POS) tagging is done, which identifies nouns, verbs, adjectives, and other parts of speech in context.
3. Using the built-in classifier provided with the

**Table 1**

Location information extracted from example tweets

| Example Tweet | Extracted Location Information |
| --- | --- |
| @MTA @NYCTSubway currently at Grand Ave/ Newtown...can you send someone?? | Grand Ave, Newtown |
| I'm at LaGuardia Airport (LGA) in East Elmhurst, NY | Laguardia Airport, East Elmhurst |
| Accident in #TheBronx on The Bronx River Pkwy SB approaching 177th St, stop and go traffic back to Boston Rd, delay of 2 mins #traffic | Bronx River Pkwy Sb, Boston Rd, 177th St |

### 4.6 Tweet location identification

The most convenient method for acquiring the geocode data from a tweet entails extracting the latitude-longitude information from the 'geo' field associated with the tweets. This field provides information on the point location where the tweet is created. Many public agencies provide real-time incident information on Twitter with the 'geo' information, where the 'geo' field resembles the incident location. After experiencing any traffic event, people can also tweet from their personal devices that are geo-tagging service-enabled. geo-enabled tweets from individuals are not very common. To overcome this limitation, location information derivation from the tweet text data is performed in this study. For example, general public posted the following tweets with specific location information: "@MTA @NYCTSubway currently at Grand Ave/ Newtown...can you send someone??" or "I'm at LaGuardia Airport (LGA) in East Elmhurst, NY." Public agencies also provide street name-embedded tweets such as "Accident in #TheBronx on The Bronx River Pkwy SB approaching 177th St, stop and go traffic back to Boston Rd, delay of 2 mins #traffic". To extract the location/street information from the tweet, the Named Entity Recognition (NER) task was performed with the NLTK module (Bird, Klein, &

NLTK module, location information is extracted from each tweet. Necessary revisions in the POS tagging task are done to accurately extract the location names. The extracted location names from the sample tweet texts are provided in Table 1.

The extracted location information is tokenized and sorted, and finally matched with the Street Name Dictionary (SND) list (DCP, 2019). Developed by the NYC Department of Planning, the SND file contains the information of the geographic features, including street names, of the entire city of New York. The match between the location names from the tweets and SND file was calculated using the similarity ratio (i.e., the closeness of two strings expressed from 0 to 100) based on the Levenshtein distance (Cohen, 2011; Occen, 2016). If $x$ (i.e., tweet) and $y$ (i.e., SND record) are two strings, and $a$ and $b$ are the length of these strings, respectively, the similarity between these strings are defined as (Cohen, 2011):

$$S_{x,y}(a, b) = \frac{2*m}{(a+b)} \leq \alpha \qquad (12)$$

where $m$ is the number of matched elements in strings $x$ and $y$, and $\alpha$ is the acceptable threshold of steps associated with this tweet coordination retrieval task are illustrated in Fig. 2.

**Table 2**
Tweet data amount per day

| Tweet Type | Number of tweets | | | | | | | Total tweets per class |
|---|---|---|---|---|---|---|---|---|
| | Monday | Tuesday | Wednesday | Thursday | Friday | Saturday | Sunday | |
| Non-transportation related tweets | 103,547 | 95,807 | 115,389 | 96,674 | 94,638 | 77,379 | 98,450 | 681,884 |
| Transportation related tweets | 2,973 | 3,125 | 2,333 | 2,997 | 3,338 | 1,931 | 1,429 | 18,126 |
| Total tweets per day | 106,520 | 98,932 | 115,391 | 99,671 | 97,976 | 79,310 | 99,879 | Total : 700,010 |

the ratio to consider a match between $x$ and $y$. Once both the on street and cross streets are identified in the SND list based on the $\alpha$, their boroughs are matched. In NYC, the same street name can often be found in different boroughs. Extracting and matching the borough names from the SDN file limits the possibility of locating the incident in the wrong borough. Using the street names and borough information, the intersection coordinate is found in the NYC geoclient API (Krauss, 2014). If no record of the intersection is found using this

## 5. Twitter data analysis

### 5.1 Tweet dataset description

Table 2 shows the amount of data collected from the case study area for each day. The initial SVM-based classification of transportation-related and non-transportation related tweets for each day are conducted using the total number of tweets collected each day. After assessing the performance of SVM, an analysis is conducted using only the transportation-related tweets (i.e., the 18,126 tweets) to evaluate the performance of

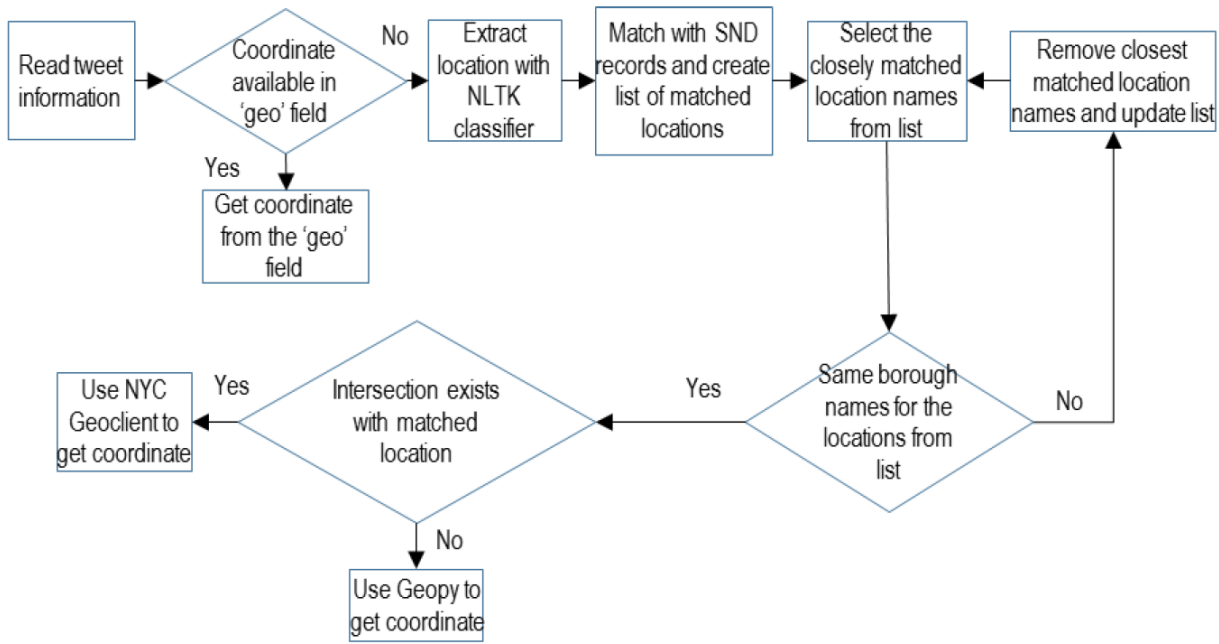

**Fig. 2.** Geoding the tweets

API, the coordinate is derived using the borough name and any one of the street names with the geopy package as suggested in (Russell, 2011). The

L-LDA, SVM, and L-LDA incorporated SVM classifiers.

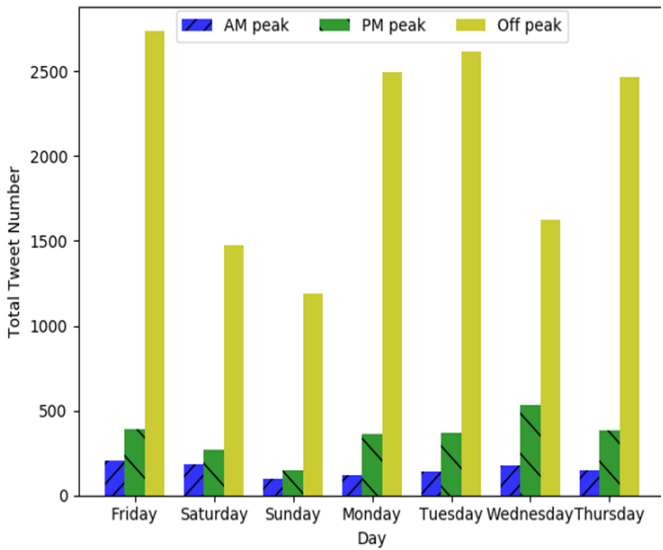

(a) Aggregated Twitter data for each day

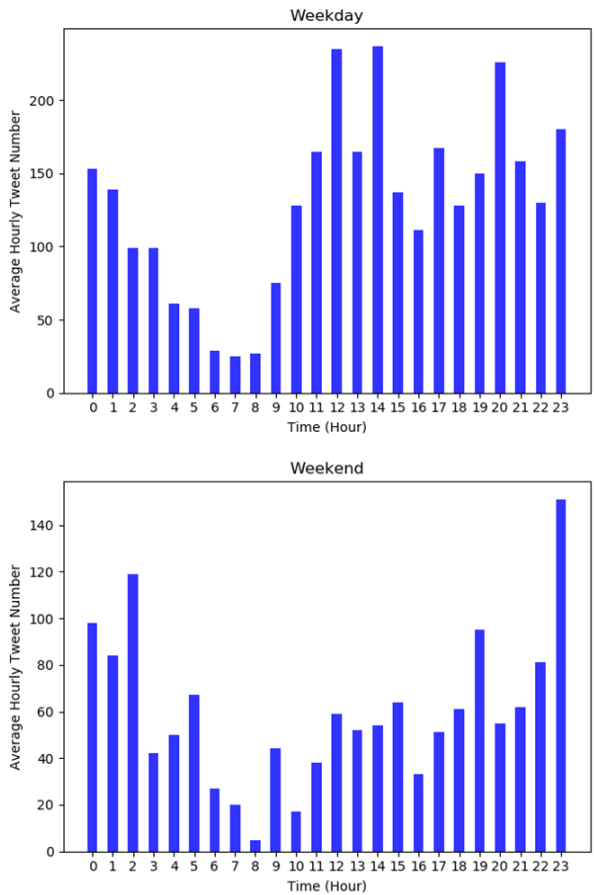

(b) Weekday and weekend average hourly tweet data

**Fig. 3.** Temporal Distribution of Twitter information.

### 5.2 Temporal distribution of Twitter data and influential users

Fig. 3 shows the distribution of Twitter information with times for (a) each day and (b) for weekday and weekend with the average value for transportation-related tweets. Fig. 3(a) shows the data for peak (morning peak: 6 am-10 am, afternoon peak: 4 pm-7 pm) and off-peak (7 pm-6 am, 10 am-4 pm) periods. Fig. 3(b) shows that on average Twitter produces more transportation-related tweets on a weekday than a weekend.

As shown in Table 3, among the transportation-related tweets, very few tweets are generated by the general public and other accounts. For the selected week, the general public mostly used Twitter while using different subways in NYC, or when they are at the airports. On Monday, Tuesday and Thursday, only 4% of the total transportation-related tweets are generated by the general public and other accounts, and on Saturday 22% of the transportation-related tweets are generated from the general public and other accounts. Table 4 demonstrates the number of tweets generated by different user groups for the transportation-related sub-classes. The main influential users are 511 and TotalTraffic, and tweets generated from these accounts have geolocation information. In Twitter, both 511 and TotalTraffic accounts in New York are specific agency-based accounts that distribute transportation-related information across New York City. The 511NY Twitter account (e.g., 511NY system) automatically distributes structured information, based on the data collected from the police department, transportation agencies, 911 calls, construction crews, motorist assistance patrol drivers, transit agencies and roadway sensors (i.e., traffic camera). The TotalTraffic account, distributes structured data based on the data collected by a private company, titled "Total Traffic and Weather Network". Instead of using the publicly available Twitter data, if the data from the Twitter Firehose (where 100% Twitter data is available) can be used, the scenario will differ given the availability of additional tweets from Twitter. However, as the Twitter Firehose is not used for this research, only publicly available Twitter data is used.

### 5.3 Feature selection for tweet classification

Based on the Lasso feature selection method, five unique numeric features are identified for SVM: sentiment score, length of a tweet, number of hashtags, number of exclamation marks, and number of question marks. This test is conducted with data from Monday. For the tf-idf vector, the dimension is reduced by T-SVD. For T-SVD, the reduced dimension of the data is assessed using cross-validation method. Using the Saturday training dataset (as it was the initial day of data collection) the accuracy of the SVM method with multinomial topic distributions over the entire vocabulary of each data.

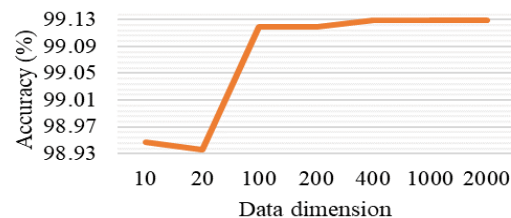

**Fig. 4.** T-SVD data dimension and corresponding accuracy

**Table 3**

Users/account holders generating transportation related tweets

| User | Monday Tweet (% of total Monday Tweet) | Tuesday Tweet (% of total Tuesday Tweet) | Wednesday Tweet (% of total Wednesday Tweet) | Thursday Tweet (% of total Thursday Tweet) | Friday Tweet (% of total Friday Tweet) | Saturday Tweet (% of total Saturday Tweet) | Sunday Tweet (% of total Sunday Tweet) |
|---|---|---|---|---|---|---|---|
| 511 (511NY, 511NYC etc.) | 2,600 (88%) | 2,704 (86%) | 1,986 (85%) | 2,665 (89%) | 2,985 (89%) | 1,405 (73%) | 1,151 (81%) |
| Total Traffic | 242 (8%) | 300 (10%) | 182 (8%) | 216 (7%) | 199 (6%) | 106 (5%) | 115 (8%) |
| General public and others | 131 (4%) | 121 (4%) | 165 (7%) | 116 (4%) | 154 (5%) | 420 (22%) | 163 (11%) |

**Table 4**

Number of transportation-related tweet per user group

| Transportation sub-class | Number of Total Tweet per User Group | | |
|---|---|---|---|
| | 511 service provider | TotalTraffic service provider | Others users |
| Construction | 3993 | 16 | 4 |
| Traffic Operations | 93 | 257 | 87 |
| Incident | 11322 | 1080 | 35 |
| Special Events | 86 | 2 | 0 |
| Others Events | 2 | 5 | 1144 |

different dimension sizes is evaluated to classify the transportation and non-transportation data. From the following Fig. 4, it is observed that after 400 and more dimensions, the accuracy of SVM classification does not improve. Based on this finding, T-SVD with 400 dimension is considered for the later analysis in this study. For L-LDA, no feature selection is needed to identify the sub-classes of the tweets since L-LDA creates the

### 5.4 Parallel computation efficacy for transportation-related tweet classification

While classifying the whole dataset with almost 700,010 tweets, SVM parameters need to be optimized, and the appropriate kernel function needs to be identified. Once the features are selected, a grid-search method is used to identify the optimal parameter for SVM. For this task, stratified sampling is used, which creates equally balanced transportation and non-transportation training dataset to find the optimal parameters, as the number of non-transportation related tweets are parameter optimization task requires, on average, around 30 minutes to execute. Using the optimized parameters a single training and validation process

**Table 5**

Transportation and non-transportation classifier accuracy (for both structured and unstructured data)

| Tweet Type | Accuracy (%) | | | | | Recall (%) | | Precision (%) | | RMSE | |
| | Tweets with structured data | | Tweets with unstructured data | Tweets with structured and unstructured data | | Tweets with structured and unstructured data | | | | | |
| | 511 service providers | TotalTraffic service providers | Other users | For each tweet type (all users) | Overall | For each tweet type (all users) | Overall (Marco-average) | For each tweet type (all users) | Overall (Marco-average) | For each tweet type (all users) | Overall |
| Non-transportation | N/A* | N/A* | N/A* | 99.9 | | 99.9 | | 99.8 | | 0.02 | |
| | | | | | 99.7 | | 95.3 | | 98.9 | | 0.053 |
| Transportation | 97.4 | 92.9 | 6.8 | 90.7 | | 90.8 | | 98.2 | | 0.3 | |

*\*User group-specific evaluation is not conducted for non-transportation data*

higher compared to the transportation-related data in the training dataset. As the classification task requires intensive computation, the Clemson requires, on average, 8 and 10 minutes to execute, respectively.

**Table 6**

Confusion matrix for SVM classifier

| | Predicted non-transportation related tweets | Predicted transportation related tweets |
| --- | --- | --- |
| Actual non-transportation related tweets | True Negative = 136,308 | False Positive = 60 |
| Actual transportation related tweets | False Negative = 335 | True Positive = 3,289 |

**Table 7**

Transportation and non-transportation classifier accuracy (for only unstructured data)

| Tweet type | Accuracy (%) | | Recall (Marco-average) (%) | Precision (Marco-average) (%) | RMSE |
| | For each tweet type | Overall | | | |
| --- | --- | --- | --- | --- | --- |
| Non-transportation | 82.9 | | | | |
| | | 83 | 68.1 | 50 | 0.4 |
| Transportation | 53.2 | | | | |

University Palmetto supercomputing cluster is used to run the testing 30 times, following the study conducted by (Singh, Lucas, Dalpatadu, & Murphy, 2013), with random training and test samples. Using parallelization, the SVM parameter optimization and classification tasks have achieved 30 times speedup compared to sequential computing. To identify the transportation and non-transportation related tweets, a single SVM

### 5.5 Performance of tweet classification
#### 5.5.1 Transportation and non-transportation event identification using all tweets

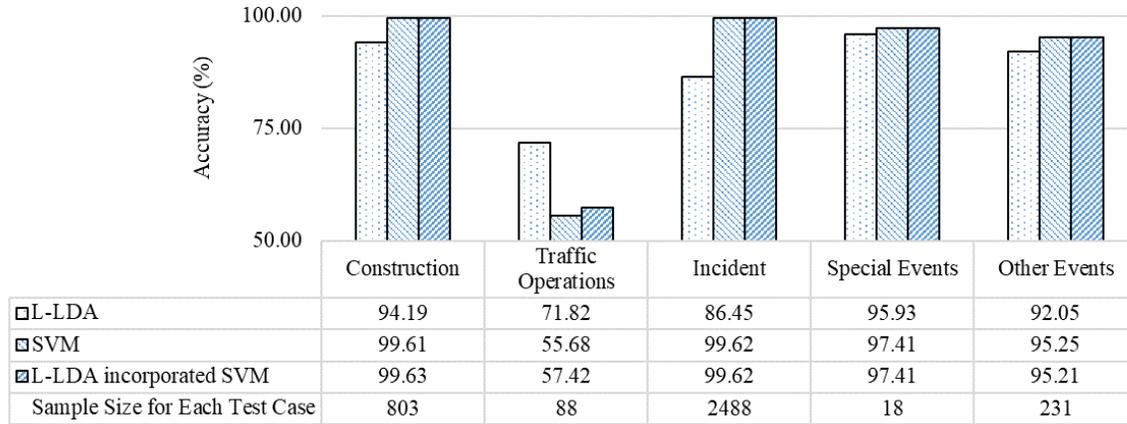

| | Construction | Traffic Operations | Incident | Special Events | Other Events |
|---|---|---|---|---|---|
| ☐ L-LDA | 94.19 | 71.82 | 86.45 | 95.93 | 92.05 |
| ☐ SVM | 99.61 | 55.68 | 99.62 | 97.41 | 95.25 |
| ☒ L-LDA incorporated SVM | 99.63 | 57.42 | 99.62 | 97.41 | 95.21 |
| Sample Size for Each Test Case | 803 | 88 | 2488 | 18 | 231 |

**Fig. 5.** Classifier average accuracy for test transportation-related tweets.

Each tweet is manually labeled to study the accuracy of supervised classifiers. After cross-validation, the linear kernel function is found to provide higher accuracy than other kernel function. Average accuracy (for running the test 30 times) of the SVM classification model is found to be 99% for each day to classify the transportation and non-transportation related tweets (including both structured data from 511 and TotalTraffic, and unstructured data from other users including the general public and news media). The accuracy of classifying transportation and non-transportation related tweets are 99.9% and 90.7%, respectively, as shown in Table 5. It also shows that machine learning-based classifier is not able to identify the unstructured data (accuracy is only 6.8%). Only

7% of the total transportation-related tweets have the unstructured format. The language used in the unstructured data is extremely diversified. Based on the findings, unstructured tweets cannot properly be classified if the classifier is developed using both structured and unstructured tweets from NYC. The precision and recall values are 98.9% and 95.3% respectively. Table 6 shows the confusion matrix of the classifier accuracy.

### 5.5.2 Transportation and non-transportation event identification using unstructured tweets

Another classification task is conducted with only unstructured data. Data from 511 and TotalTraffic

**Table 8**
Classifier accuracy for transportation-related sub-classes by different classifiers

| Classifier | Accuracy (%) | | | | Recall (Marco-average) (%) | Precision (Marco-average) (%) |
|---|---|---|---|---|---|---|
| | Tweets from users with structured data | | Tweets from users with unstructured data | Tweets from users with both structured and unstructured data | Tweets from users with both structured and unstructured data | Tweets from users with both structured and unstructured data |
| | 511 Service Provider | TotalTraffic Service Provider | Other Users | Overall | Overall | Overall |
| L-LDA | 91.8 | 49.4 | 85.9 | 88.2 | 88.1 | 64.3 |
| SVM | 99.4 | 94.4 | 88.7 | 98.2 | 89.5 | 93.4 |
| L-LDA incorporated SVM | 99.4 | 95.2 | 88.5 | 98.3 | 90 | 94.4 |

**Table 9**

Evaluation of L-LDA incorporated SVM (percentages are shown in parenthesis)

| Actual Class | Predicted Class Sample size (Classification/misclassification accuracy %) | | | | |
|---|---|---|---|---|---|
| | Construction | Traffic Operations | Incident | Special Events | Other Events |
| Construction | 800 (99.5) | 2 (0.25) | 1 (0.12) | 0 (0.0) | 1 (0.12) |
| Traffic Operations | 1 (1.12) | 51 (57.3) | 25 (28.1) | 0 (0.0) | 12 (13.5) |
| Incident | 0 (0.0) | 3 (0.12) | 2479 (99.64) | 0 (0.0) | 6 (0.24) |
| Special Events | 0 (0.0) | 0 (0.0) | 0 (0.0) | 18 (100) | 0 (0.0) |
| Other Events | 1 (0.43) | 5 (2.16) | 5 (2.16) | 0 (0.0) | 220 (95.24) |

**Table 10**

Precision and recall for L-LDA incorporated SVM

| Measures | Sub-class | | | | |
|---|---|---|---|---|---|
| | Construction | Traffic Operations | Incident | Special Events | Other Events |
| Precision | 99.8% | 83.3% | 98.8% | 98.1% | 92.2% |
| Recall | 99.6% | 57.42% | 99.6% | 97.4% | 95.2% |

**Table 11**

L-LDA identified top words for transportation sub-classes

| Transportation sub-class | L-LDA-identified top words |
|---|---|
| Construction | street, north, cleared, station, both, url, new, update, west, exit, wb, eb, sb, nb, construction, directions, avenue |
| Traffic Operations | closed, eb, traffic, minutes, closure, path, train, nyc, both, avenue, restrictions, new, side, update, ave, delay, queens, sb, ramp, nb, directions, url, wb, services |
| Incident | street, incident, traffic, cleared, station, both, url, new, update, exit, wb, eb, expressway, sb, nb, directions, avenue |
| Special Events | highway, special, event, plaza, sb, update, wb, service, center, side, traffic, toll, bound, cleared, streets, both, url, level, eb, parkway, york, nb, construction, broadway, avenue, interchange, east |
| Others | traffic, bus, ny, train, uber, york, new, terminal, my, airport, driver, car, subway, url, flight, mta, nyc |

accounts are excluded. As the number of transportation-related tweets is relatively small than the number of non-transportation tweets, the random under-sampling method (Galar, Fernandez, Barrenechea, Bustince, & Herrera, 2012) is used to train the classifier. In the random under-sampling method, the sample distribution for different classes is balanced by the random elimination of the samples from the class with a higher sample size. For each evaluation, a total number of 1016 non-transportation and transportation-related tweets are used to train the SVM classifier. In the test cases, 680,868 non-transportation and 254 transportation-related tweets are used. The SVM classifier parameters

(i.e., C and gamma) for this step using cross-validation. The study revealed that the radial-basis kernel function gives the highest accuracy with C=0.5, gamma=0.5. Using these values, Table 7 shows the SVM classifier accuracy using only unstructured data. The overall accuracy of the classifier using unstructured data is 83%, while for transportation-related data, it is 53.2% for 30 test cases.

### 5.5.3 Transportation events identification using all tweets

After the transportation-related tweets are extracted with SVM, three supervised classifiers

(i.e., L-LDA, SVM and L-LDA incorporated SVM) are applied to further categorize transportation-related tweets into sub-classes using both structured (i.e., data from 511 and TotalTraffic) and unstructured data (i.e., data from the general public and other news media). While compared with the manually coded labels, as indicated in Fig. 5 and Table 8, L-LDA achieves the minimum average accuracy (88.2% for 30 random tests) to classify the tweets into five sub-classes, whereas the L-LDA incorporated SVM achieves the maximum average accuracy (98.3% for 30 random tests) for the same classification. At a 90% confidence level, based on the Wilcoxon Signed Rank test, the median of the accuracy achieved by L-LDA incorporate SVM is significantly higher than the accuracy of both L-LDA and SVM for classifying transportation-related tweets. Fig. 5 also shows the average accuracy for 30 random tests of five sub-classes, and the number of tweets per sub-class for each sample. For construction and incident sub-classes, more data are available compared to other sub-classes, consequently all three classifiers achieve higher accuracy to classify tweets compared to the minimum required accuracy of 85%. L-LDA incorporated SVM achieves higher accuracy than both L-LDA and SVM classifiers for classifying in all five sub-classes, except other events where SVM achieves 0.4% higher accuracy compared to L-LDA incorporated SVM.

Table 9 shows the actual class and predicted class matrix of L-LDA incorporated SVM. The values in parenthesis show the classification/misclassification accuracy of each predicted class. It shows that for 'traffic operations', 28.1% of tweets are misclassified as 'incident'. Due to the similarity of the tweet information (i.e., roadway condition status, road blockage, clearance information, etc.) between these sub-classes, the misclassification occurs.

Table 10 shows the precision and recall values of the L-LDA incorporated SVM classifiers. It shows that tweets related to other sub-classes are not classified as 'construction' and 'incident' sub-classes, as the precision value of these two sub-classes is almost close to 100%. Based on the recall values, tweets from 'construction', 'incident', 'special events' and 'other events' sub-classes are grouped most accurately (i.e., recall value greater than 90%). Table 11 shows the top words identified by L-LDA for each transportation-related sub-class.

### 5.6 Geocoder Accuracy Analysis

Using the geocoders, coordinates of tweets records are estimated. Location names from tweets are matched with the SND dataset using the similarity ratio calculated with the Eq. 12. After cross-validation, it is found that location names are similar with similarity ratios ($\alpha$) 80 or more. For this research, $\alpha$ is taken as more than or equal to 80. Geo-enabled tweets with embedded latitude-longitude information (i.e., latitude-longitude provided in the 'geo' field) are tested to validate the performance of the geocoders. On the other hand, using a geocoder, the locations of the tweets are identified based on information from the tweet text. The geocoder-derived latitude and longitude are matched with the latitude-longitude information provided in the tweet 'geo' field. The mean value of the distance difference between latitude-longitude provided in the 'geo' field and geocoder information from the tweet text is 7.3 miles, with 8.7 miles of standard deviation. As shown in Fig. 6, the 25, 50, and 75th percentile values are 0.5, 3.9 and 10.6 miles respectively.

Using the geocoder, the tweet location for the general public is also assessed. Geolocation discrepancy exists in Twitter because motorists often mention neither street nor location names when tweeting about traffic congestion, incidents or any other events. For example, "Our Lyft driver just told me that she's only been driving for 10 days. #jesustakethewheel" or "U gotta thank the bus drivers for getting u to ya destination safe ,I really be appreciating that" are examples of transportation-related tweets which do not have any content to derive any specific location. Often the tweet is about locations out of NYC, which also did not help to generate the location of the tweet in NYC. Using geocoder, latitude-longitude information is successfully captured if the location-related text is provided in the tweet text. For example, "495 westbound out of Lincoln Tunnel is apparently closed. Thanks, #sarcasm #farehikesforwhat" or "On Atlantic Ave this morning thanks to my Brad. uber from Massachusetts? @nyctaxi" tweets have location

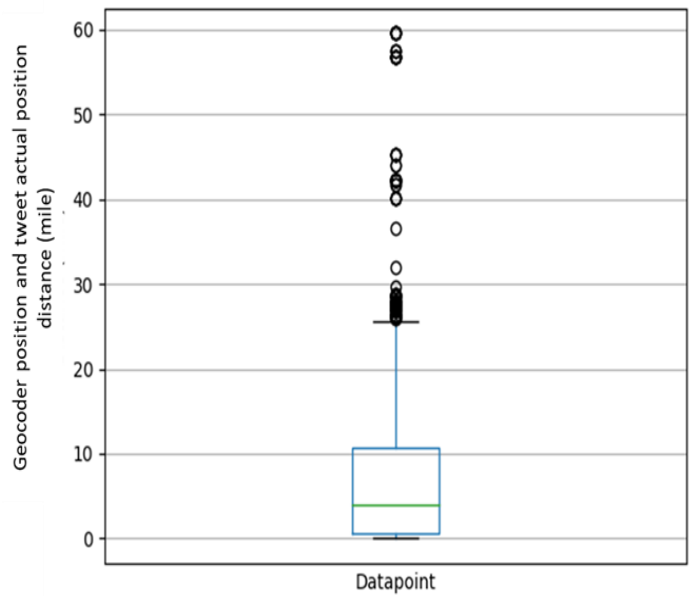

**Fig. 6.** Box plot of the geocoder position and tweet actual position distance

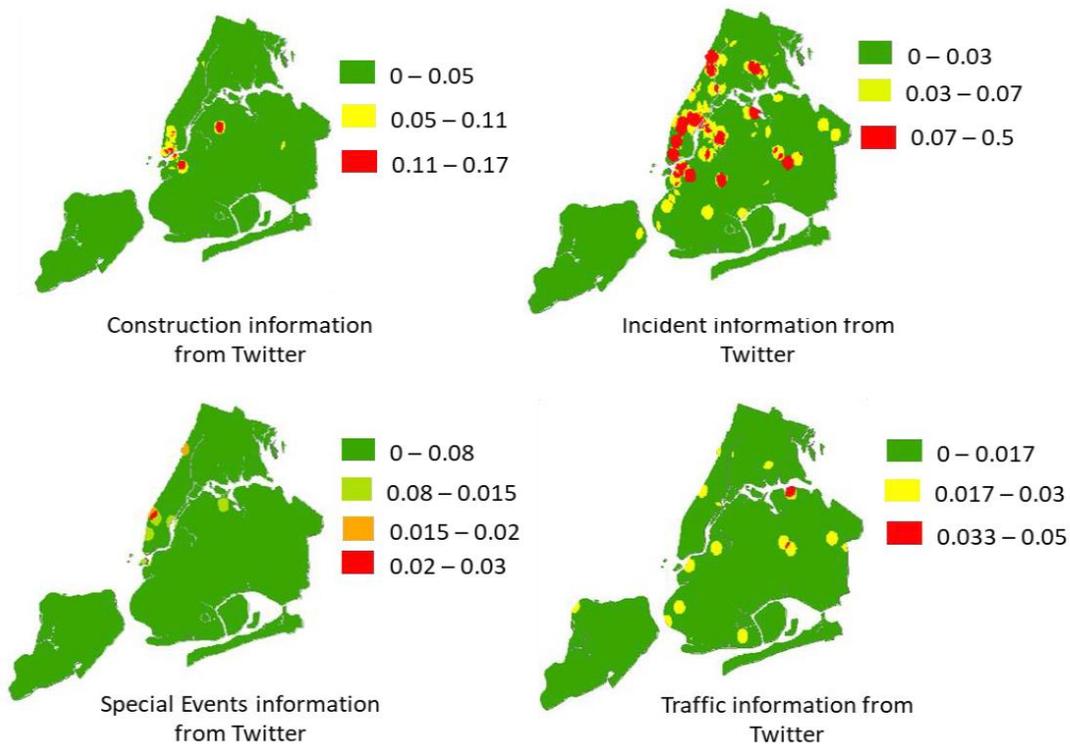

**Fig. 7.** Density map for Monday using Twitter

information, and the geocoder has successfully captured the location from the tweet text. The geocoder derived locations for these two tweets are "Lincoln Tunnel (40.7588352, -73.9999574)" and "Atlantic Avenue (40.59116, -74.090754)." Sometimes some landmarks are used in the tweet text, which also helps to identify the tweet location. For example, "Really @Uber $150 to get from JFK

to UWS? I'd say it's highway robbery but it's really more Van Wyck Robbery." has JFK airport in the tweet text, which implies that the tweet is originated from the JFK airport. The geocoder also captures such events. Once the transportation-related tweets are classified and geocoded, these tweets are projected onto a map of NYC. For example, Fig. 7 shows a density map, as a case study, using Twitter data for Monday, which has the maximum number of transportation tweets among all days of the week analyzed in this study. It shows the information gathered from Twitter for each 100 sq. ft. area for the sub-classes. It is evident from Fig. 7 that using the publicly available tweets, classified transportation-related events, such as construction, incident, special events, traffic condition, could also be captured in NYC. This suggests that Twitter can potentially provide more details about transportation-related events including the type of events.

## 6. Discussion of the results

This study has identified Twitter as a viable source of collecting transportation data by analyzing both structured and unstructured tweets. In the unstructured tweets generated by public, ambiguities exist in tweet texts, which influence the classifier performance. Also, the lack of transportation domain-specific ontology, language-related challenges (e.g., jargons in the language), imbalanced data in different classes, improper annotation, and lack of location information are some of the critical challenges to analyze the unstructured tweets (Grant-Muller et al., 2015; Kuflik et al., 2017). The general analytical framework has two steps: tweet classification and tweet geocoding. Of these steps, the tweet classification framework is transferred to other locations once the tweets are collected from those regions, and the classifiers are trained with data (both structured and unstructured) generated specifically from those regions. For tweet geocoding, the NYC geoclient and Geopy geocoders are used. The NYC geoclient is an API, which is available for NYC only. To identify locations for other areas, area-specific geocoders can be used. Also, the Geopy geocoder can be used to identify the location from any region on this planet. In the future, other publicly available databases can be augmented with the Twitter-based transportation event identification system. Publicly available navigation tools, such as Waze and Google maps, provide data on incident, construction, and major events. Google map shows the live traffic data based on historical data as well as real-time smart-phone based crowdsourced data.

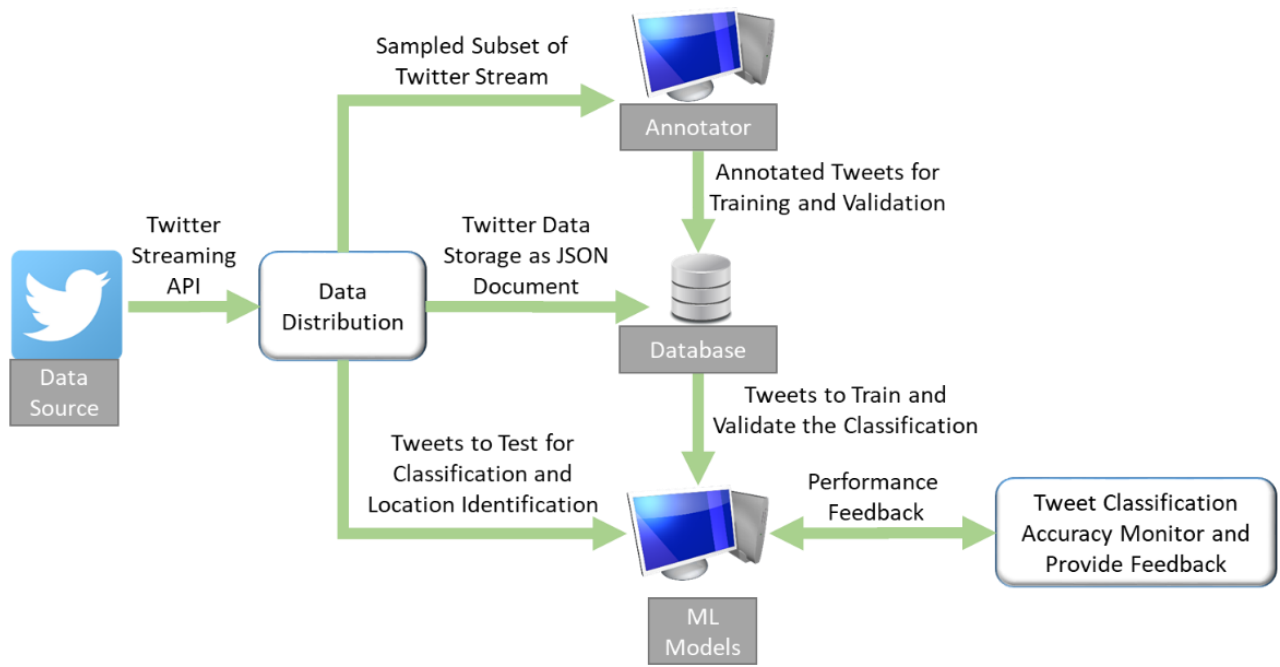

**Fig. 8.** Real-time, data-driven and feedback-based tweet identification framework.

Additional data, such as data related to construction, incident, special events, are derived from the Waze application, which is a crowdsourced-based application. Waze is a specialized social network tool for navigation, which provides customized routes for users based on the users' preferences (the route through a low-price gas station, police activities along the routes, etc.). Also, labeling of 700K tweets with multiple human annotators is a challenging and time-consuming task. In a parallel study, manual annotation of Tweet dataset on UberPool is conducted with a small sample size (i.e., 1000 tweets), where three different annotators were involved (Pratt et al., 2019). This study reported the final labels from these three annotators based on the weighted average method. The inter-annotator agreement from the parallel study on the UberPool dataset is 98% for the speaker (i.e., identifying who generated that tweet), 86% for the subject (i.e., identifying what is the tweet about), and 96% for the sentiment (i.e., identifying general emotion of the tweet). Sometimes the inter-annotator agreement rates can be low if experts do not do the annotation. In one study (Nowak & Rüger, 2010), authors evauated crowd-sourcing based annotation and found that that the non-expert based inter-annotator rate is much lower than that of the expert-based. Further, because of language ambiguity, typos, and lack of context in the texts, the inter-annotator agreement can be low. In the future, manual labeling can be performed to verify the performance of the machine-learning based auto-annotator, as shown in Fig. 8.

To increase the classifier (i.e., classifier to categorize transportation and non-transportation data) accuracy with only unstructured data, a data-driven feedback loop can be used in the framework which will monitor the performance of the machine-learning based classifier and update the database in real-time. In the unstructured tweets, general people and different news media provided information about constructions, incidents, and traffic operations. They also provided: (a) comments about the public transit, and ridesharing services, (b) update from multi-modal terminals (i.e., airport, public transit), (c) opinions about transportation events, (d) comments about other road travelers' behavior, etc. Due to the large topic variation for a single dataset having a low sample size, classification of the unstructured data from general people and news media is inherently challenging. If more data can be collected from the general public, the data can be used for crash data validation, secondary crash identification, and bottleneck extent identification due to congestion and/or construction. Using more unstructured transportation-related data in the future, the better classifier can be developed to classify the unstructured data more accurately. In one study (Holzinger, 2016), the author discussed reinforcement learning and preference learning methods to provide feedback to the machine learning models, which can be used in future for the Twitter classification framework as shown in Fig. 8. In this framework, the auto annotator will assign labels to the new training and validation tweet set, and store the data in the database. The machine learning (ML) based classifier will assign labels to the test data, which will be evaluated by the classification-performance monitoring module. Later the monitoring module will provide performance feedback to the classifier so that the ML can be updated with time to achieve better classification accuracy.

## 7. Contributions of the research

In this research, n-grams tweet classification and a geocoding framework have been developed for the real-time tweet stream classification and location identification for any region. This study fills the gap in text stream analysis by developing a tweet stream analysis framework, which is absent from the literature (Mirończuk & Protasiewicz, 2018). The framework has two components, which are tweet classification and location identification. In the tweet classification, one hybrid method is tested (L-LDA incorporated SVM) to classify the transportation-related tweets. In one study, the authors (K. Dalal & A. Zaveri, 2011) discussed the importance of developing a hybrid method for text classification to achieve better classification results. In the location identification, a novel method of identifying tweet location based on string similarity is developed, which identifies the location of tweets within 7.3 miles of the exact location within NYC. This geocoder can accurately identify locations from tweet text generated by the general public if they mention any landmark or street names within the tweet text. Using this framework, real-time tweet stream is automatically analyzed for any large region, and the extracted information used by both public and private agencies, researchers, and the general public.

## 8. Conclusions

Public or government agencies, such as transportation agency and law enforcement agency, and private companies collect, process, and disseminate traffic information as part of their services to travelers. Any accurate publicly accessible information would increase the reliability of the public or private agency collected data. Among other external data sources, the emergence of social media platforms over the last decade has created a unique platform for public agencies to collect real-time incident status information from those users with minimum resource investment. In this research, an analytical approach is developed for supporting tweet classification and string similarity based geocoding in a parallel computing environment. Once the classifiers are developed, streaming data from Twitter can be classified in real-time to identify the transportation related tweets. Developing supervised learning based classifiers for a large region using tweets is computationally expensive, as the computation time can be very high. In this research, parallel computation-enabled relevant natural language processing steps and a novel geocoding procedure have been used to overcome the inherent ambiguities of tweets and tweet analysis to extract relevant transportation-related information.

A new supervised classifier is developed so that SVM may use topic distribution probability using L-LDA into the SVM feature space. The accuracy of this classifier (i.e., L-LDA incorporated SVM) is found to be significantly higher than the accuracies of both standalone L-LDA or SVM at a 90% confidence level. The achieved 99% classification accuracy (i.e., compared to the manually coded labels) is above the minimum accuracy requirement (i.e., 85%) for the statewide incident reporting system according to the Title 23 of the Code of Federal Regulations. It is observed that 511 and TotalTraffic are the influential users of Twitter in NYC and its surrounding areas, and apart from these accounts, very limited transportation related-tweets are generated from general public and news accounts.

Also, in this research, the geo-coordinates assigned by the string-similarity based geocoding process is validated using the tweets which have geo-coordinate information available from Twitter. On average, the assigned coordinates fall within 7.3 miles of the actual tweet location. Using these accurately classified and geocoded tweets, the transportation-related information available from the general public in Twitter can be used to augment public agency collected data, such as incident data collected by the New York Police Department. Various information (e.g., incident impact, congestion extent, emergency weather) are available from Twitter, which can provide additional information to public agencies about any traffic events. The L-LDA incorporated SVM classifier can be utilized in a traffic management center or TMC to extract transportation data from publicly available tweet dataset to help manage traffic in real-time. Data from the general public can help receiving the real-time update during emergency evacuation events, or special occasions, which can help real-time traffic management as well as future traffic planning. Analysis of both structured and unstructured tweets demonstrates the feasibility of using Twitter as a viable source for transportation data collection. This research is conducted with publicly available Twitter data which is only 1% of the total Twitter dataset. If Twitter Firehose data is included, it will provide more coverage to validate the traditional roadway traffic sensor (e.g., loop detector, video camera) collected data.